\documentclass[twocolumn,showpacs,preprintnumbers,amssymb]{revtex4}
\usepackage{graphicx}
\usepackage{dcolumn}
\usepackage{bm}
\usepackage{dsfont}
\usepackage{amsmath}
\usepackage{pifont}
\usepackage{psfrag}
\usepackage[latin2]{inputenc}
\usepackage{bbm}
\usepackage{color}

\newcommand\ket[1]{|{#1}\rangle}

\newcommand{\BE}{\begin{equation}}
\newcommand{\EE}{\end{equation}}

\newcommand{\skipc}[2]{}

\newcommand{\fig}[1]{Fig.~\ref{#1}}
\newcommand{\eq}[1]{Eq.~(\ref{#1})}

\newcommand{\I}{\ensuremath{{\mkern1mu\mathrm{i}\mkern1mu}}}
\newcommand{\E}{\ensuremath{{\mkern1mu\mathrm{e}\mkern1mu}}}

\newcommand{\qed}{\nobreak \ifvmode \relax \else
      \ifdim\lastskip<1.5em \hskip-\lastskip
      \hskip1.5em plus0em minus0.5em \fi \nobreak
     $\square$\fi}

\begin{document}

\preprint{APS/123-QED}

\title{Quantum dynamics of trapped ions in a dynamic field gradient using dressed states} 
\author{Sabine W\"olk and Christof Wunderlich}

\affiliation{Department Physik, Naturwissenschaftlich-Technische Fakult\"at, Universit\"at Siegen, 57068 Siegen, Germany }

\date{June 9, 2016}

\begin{abstract}

Novel ion traps that provide either a static or a dynamic magnetic gradient field allow for the use of 
radio frequency (rf) radiation for coupling internal and motional states of ions, which is essential for conditional quantum logic. 
We show that the coupling mechanism in the presence of a dynamic gradient is the same, in a dressed state basis, as in the case of a static gradient. Then, it is shown how demanding experimental requirements arising when using a dynamic gradient could be overcome.
Thus, using dressed states in a dynamic gradient field could decisively reduce experimental complexity on the route towards a scalable device for quantum information science based on rf-driven trapped ions.     

\end{abstract}



\maketitle

Experiments with atomic trapped ions have played a leading role in the development of experimental quantum information science \cite{Cirac1995,Blatt2008,Wineland2013}.
Well isolated from their environment, trapped ions are ideally suited for investigating fundamental questions of quantum physics, and are a promising candidate for quantum simulations and scalable universal quantum computing reaching  beyond the capabilities of classical computers \cite{Monroe2010}.
Internal electronic states serving as qubits are coherently prepared using electromagnetic radiation in the optical or radio-frequency (RF) regime,  and an upper limit for the coherence time of ionic qubits is set by the coherence time of this radiation. 
For conditional quantum dynamics with two or more qubits, represented by several ions confined in the same trap or trapping region, the collective vibrational motion is coupled to the internal dynamics of individual ions, thus serving as a quantum bus. 

Using laser light for coupling ionic qubits via this quantum bus 
has been standard for some decades, since only with light in and around the visible regime the Lamb-Dicke parameter $\eta$, measuring the coupling strength between internal and motional states \cite{Stenholm1986}, takes on a sufficiently large value in typical traps. Driving solely a single desired ion out of a collection of trapped ions, typically spaced apart by a few micrometers, also required optical radiation that can be focused down to a spot size smaller than the inter-ion separation.
In numerous experiments laser light has been successfully used to deterministically prepare  quantum states of trapped ions, even complete quantum algorithms \cite{Hanneke2010,Blatt2016} and quantum simulations \cite{Schneider2012,Islam2013} have been implemented. 

The complexity of experimental set-ups 
can be reduced decisively, when RF radiation is used  to directly drive the ions' dynamics instead of taking the detour of imprinting RF signals onto optical beams and then steering these optical beams towards trapped ions. With laser beams, frequency-, phase-, and amplitude noise, diffraction, and beam pointing instabilities in the optical domain pose additional problems that can be avoided by the direct use of RF radiation. 

Using  RF radiation for coupling internal and motional dynamics becomes possible when an additional, spatially varying field is applied to an atom trap. This can be a static \cite{Mintert2001} or a dynamic \cite{Ospelkaus2008} magnetic gradient field. In both cases, an effective Lamb-Dicke parameter arises through magnetic gradient induced coupling (MAGIC) even upon excitation with RF radiation \cite{Johanning2009,Ospelkaus2011,Khromova2012,Belmechri2013,Lake2015,Weidt2016}.  In addition, individual addressing of atoms using RF radiation has been shown to be effective \cite{Schrader2004,Johanning2009,Khromova2012,Warring2013,Piltz2014,Lake2015,Aude2016}. 

When employing MAGIC for trapped ions as a complimentary approach to successful research based on laser-driven ion trap quantum logic,  spontaneous emission because of the finite lifetime of qubit states, or spontaneous scattering caused by non-resonant laser light driving Raman transitions is not a concern for the coherence time of qubits. Also, RF-based single- and multi-qubit gates can be tolerant against thermal excitation of the ions' vibrational motion. 

Single-qubit quantum gates driven by RF radiation have been implemented with an error well below $10^{-4}$ \cite{Brown2011,Lucas2014}, an important threshold for fault-tolerant quantum computing. 
Using a static field gradient, a quantum byte (eight ions) could be addressed with a measured cross-talk between closely spaced, interacting ions in the $10^{-5}$ range \cite{Piltz2014}.  MAGIC was also employed to demonstrate  two-qubit gates \cite{Ospelkaus2011,Khromova2012,Weidt2016}, three-qubit gates \cite{Piltz2015}, and opens new possibilities for quantum simulations and quantum computation \cite{Johanning2009b,Piltz2015}.

In this Letter we show that the addition of either a static or a dynamic gradient field to a Coulomb crystal of trapped ions -- in order to take advantage of MAGIC -- can be described in an equivalent way. It is shown that the Hamiltonian in a dressed state picture, obtained when applying a spatially varying dynamic field is identical to the case of having a static gradient field and a spatially constant qubit driving field. 

In current experiments where a dynamic gradient field is applied, great care is taken to null the dynamic magnetic field at the ions' positions and thus to retain only a gradient of the dynamic field at this position \cite{Ospelkaus2011,Warring2013b,Carsjens2014} in order to obtain high fidelity two-qubit gates.  Here, we show how gates using dressed states in a dynamic magnetic gradient could be used with a non-zero  field at the ions' location, thus considerably simplifying the experimental effort  necessary when using the dynamic MAGIC scheme. 

Also, atomic states dressed by the dynamic gradient field are insensitive to ambient field noise making it superfluous to apply  a relatively strong and stable bias field in order to create qubits with long coherence time. In addition, applying a dynamic field with an amplitude gradient along the axis of weakest confinement of ions in a linear trap becomes feasible, and, thus can enhance the coupling strength between qubit states and motional states. Furthermore, it is shown that long range spin-spin coupling between dressed qubits arises useful for quantum simulations and computation.  Moreover, it is shown how ions' exposed to such a dynamic field gradient could be addressed individually. 

In what follows, we consider coupling between internal and motional states in the presence of a static or dynamic gradient field. 
We recapitulate both methods and bring them into a common representation starting by first considering a single atom before demonstrating the equivalence for multi-qubit systems.

{\em Static magnetic gradient}
For 
a static magnetic gradient parallel to the z-axis, 
the Hamiltonian describing a single atom  with energy level spacing $\hbar \omega_0$ at position $z$ expanded up to first order in the field gradient is given by 
\BE
H_\text{static}= \frac{1}{2}\hbar\omega_0 \sigma_z+\frac{\mu}{2}(B_0+zB')\sigma_z +  \hbar\nu_n   a^\dagger_n a_n,
\label{eq:H_static_1}
\EE
with the atom's magnetic dipole moment $\mu$, the magnetic field $B(z)=B_0+zB'$, and the Pauli-z matrix $\sigma_z$. Here, $a^\dagger_n$ and $a_n$ describe the creation and annihilation operators of the vibrational mode with frequency $\nu_n$. The displacement  $\Delta z$ of the ion $j$ from its equilibrium position $z_j$ can be written in terms of the normal vibrational mode $n$
\BE
\Delta z=  b_{j,n} q_n(a_n+a_n^\dagger)\label{eq:deltaz}
\EE
with the help of the expansion coefficients $b_{j,n}$ (if only a single ion  is considered, then $b_{1,1}\equiv 1$). Here, 
$q_n=\sqrt{\hbar/(2m\nu_n)}$ describes the atoms spatial extend. As a consequence, Hamiltonian \eq{eq:H_static_1} can be rewritten as
\BE
\begin{split}
 H_\text{static}=&\frac{1}{2}\hbar \omega(z_j)\sigma_z^{(j)}+ \hbar\nu_n  a^\dagger_n a_n\\
 &+\hbar \nu_n \varepsilon_{j,n} (a^\dagger_n +  a_n)\sigma_z^{(j)} \label{eq:H_static}
\end{split}
\EE
with the position dependent level splitting  $\omega(z_j)=\omega_0+\mu(B_0+z_jB')/\hbar$ and the coupling strength 
\BE
\varepsilon_{j,n}=(\mu B' b_{j,n} q_n)/(2\hbar \nu_n) \  .
\EE

The interaction between a harmonically trapped atom exposed to a static magnetic field gradient (as set out above) and an additional electromagnetic field that drives an atomic resonance is described by the Hamiltonian
\BE
\tilde{H}_D = \frac{\hbar \Omega_D}{2} \left(\tilde{\sigma}_+ \E^{\varepsilon_{j,n}(\tilde{a}^\dagger -\tilde{a})}+\tilde{\sigma}_- \E^{-\varepsilon_{j,n}(\tilde{a}^\dagger -\tilde{a})} \right) \ ,
\label{eq:H_D}
\EE
where we have neglected terms involving the Lamb-Dicke parameter $\eta =b_{j,n}q_n k$, 
since for RF radiation,  $\eta$  in usual ion traps is negligibly small. 
The Hamiltonian $\tilde{H}_D$ is obtained from the usual trapped atom-radiation interaction Hamiltonian $H_D$ after applying the  unitary transformation $\tilde{H}_D= \E^S H_D \E^{-S}$ with $S=\varepsilon_{j,n}(a_n^\dagger- a_n)\sigma_z$  \cite{Mintert2001}. 

This transformation also reveals direct interaction of the internal degrees of freedom in the case of $N$ ions confined in a linear trap described by a generalization of \eq{eq:H_static_1}. As a consequence, the static magnetic gradient induces a long range interaction \cite{Wunderlich2002,Wunderlich2003} \BE
H_{J}=-\frac{\hbar}{2} \sum\limits_{j<k} J_{j,k} \sigma_z^{(j)}\sigma_z^{(k)}
\EE
between the ions' internal states  (henceforth referred to as spins)
with the coupling strength given by
\BE
J_{j,k}=\sum\limits_n \nu_n \varepsilon_{j,n} \varepsilon_{k,n}
\label{eq:J}.
\EE
This  spin-spin coupling is up to second order independent of the motional degree of freedom and enables therefore so called hot quantum gates. 

{\em Dynamic magnetic gradient}
The scheme described in \cite{Ospelkaus2008} allows for spin-spin coupling via a dynamic magnetic field with a gradient of the amplitude perpendicular to the string of ions. When recapitulating this scheme, we take the gradient of the magnetic field $B(z,t)=\cos(\omega_B t)B(z)$ to point along the z-axis. Thus, the string of ions is parallel to the x-axis and parallel to the electrode providing the oscillating magnetic fields. 
In analogy to the static case, we express the position $z$ via the equilibrium position  $z_j$ plus a small displacement $\Delta z$ and expand the magnetic field  $B(z)=B_j+\Delta z B'$ around $z_j$. In this way, we arrive at the Hamiltonian of a single ion $j$ coupled to the radial mode $n$  
\BE
\begin{split}
H_\text{osci}= &\frac{\hbar \omega_0}{2} \sigma_z^{(j)} + \hbar \nu_n a^\dagger_n a_n+\sigma_x^{(j)} \cos(\omega_B t)\mu B_j\\
&+ \sigma_x^{(j)} \cos(\omega_B t)\mu B'\left[b_{j,n}q_n (a^\dagger_n +  a_n)\right].\label{eq:H_osci}
\end{split}
\EE
As a consequence, the  interaction Hamiltonian in the rotating wave approximation is given by
\BE
H_\text{osci,I}=-\hbar \sigma_+ (\Omega_0 e^{-\I\delta_0 t}+  \Omega_{n,j}  a_n e^{-\I(\delta_0+\nu_n) t})+\text{h.c.}
\label{eq:H_osc_I}
\EE
with the detuning $\delta_0=\omega_B-\omega_0$, the Rabi frequencies 
\BE
\Omega_0=B_j \mu /(2\hbar) \ ,
\label{eq:Omega_0}
\EE and  
\BE
\Omega_{n,j}=B' \mu b_{j,n}q_n/(2\hbar).
\label{eq:Omega_nj}
\EE
Similar to the static case, the coupling between internal and motional degrees of freedom is caused by the magnetic gradient (Eqs. \ref{eq:H_osci} and \ref{eq:H_osc_I}). 

By applying two dynamic magnetic fields with equal amplitude and opposite detuning  close to the  red- and blue sideband,  a spin-spin interaction can be generated \cite{Ospelkaus2008}.  In general, a spin-spin interaction without additional excitations via the carrier transition is desirable.  Excitation via the carrier is suppressed due to off-resonant excitation (detuning of the driving fields by  about $\nu_n$), and it can be neglected if $\Omega_0\ll \nu_n$. This leads, together with \eq{eq:Omega_0}, to the restriction  $\mu B_j\ll \hbar \nu_n$ for the magnetic field. Typical magnitudes are $\nu_n=2\pi\times10^6$ Hz
and $\hbar/\mu_B=10^{-10}$ T/Hz which leads to $B_j\ll 10^{-4}$ T. As a consequence, a high magnetic gradient and a small absolute magnetic field strength are needed for high  fidelity  two-qubit gates. Therefore, considerable experimental effort is devoted to an exact geometry of the electrodes generating the magnetic fields and to exact positioning of the ions \cite{Ospelkaus2011,Warring2013b,Carsjens2014}.



\begin{figure}
\begin{center}
\includegraphics[width=0.45\textwidth]{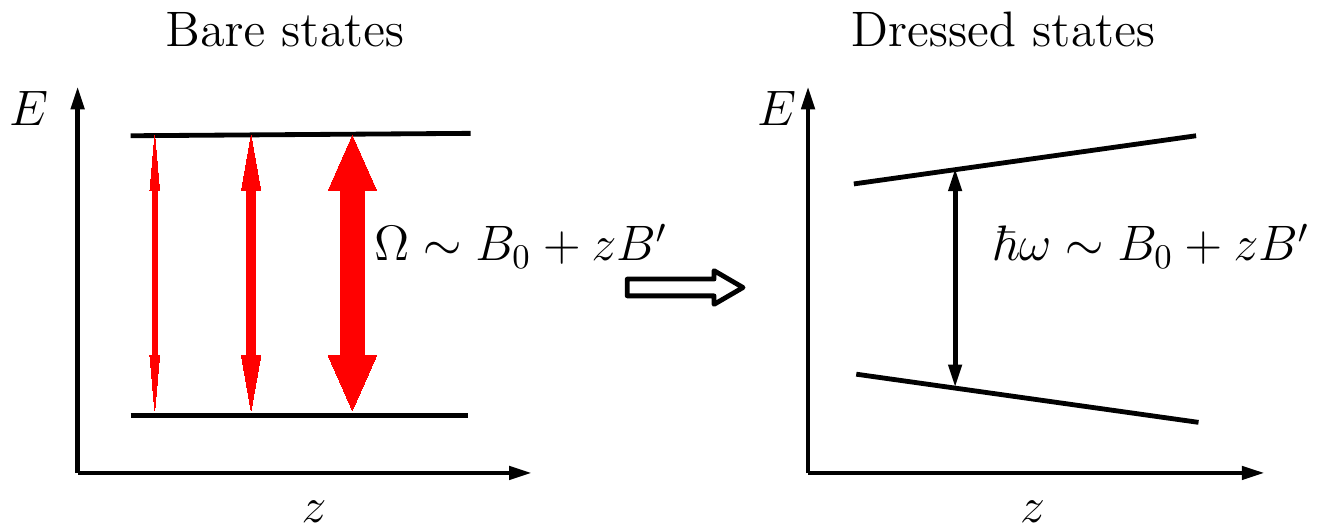}
\end{center}
\caption{Illustration of the transformation of a position dependent strength of the driving field in a dynamic gradient (left) into a position dependent level splitting of a dressed-state qubit (right)}\label{fig:dressed}
\end{figure}

{\em Equivalence of static and magnetic gradient.}
In what follows we show that the approach using a dynamic magnetic gradient is equivalent to the static gradient approach when using dressed states. For this purpose, we transform the Hamiltonian $H_\text{osci}$ given in \eq{eq:H_osci} into the rotating frame of the ion, resulting in 
\BE
\begin{split}
H_I= &\frac{\mu}{2} (\sigma_+\E^{-\I\delta t}+\sigma_-\E^{\I\delta t})\\
&\times \left\lbrace B_0+B'\left[z_n+q_j b_{j,n} (a_j^\dagger +  a_j)\right]\right\rbrace
\end{split}
\EE
for a single ion and a single mode.

Now, we further transform this interaction Hamiltonian into the dressed state picture, were the dressed states are defined in terms of the bare atomic states $\ket{g}$ and $\ket{e}$: 
\BE
\ket{\pm}=\frac{1}{\sqrt{2}}(\E^{\I\delta t/2}\ket{g}\pm \E^{-\I\delta t/2}\ket{e}).
\EE
 As a consequence, the Pauli-z matrix in the dressed state picture is given by $
\sigma_z^{(\pm)}=\sigma_+\E^{-\I\delta t}+\sigma_-\E^{\I\delta t}.$
This relation points out that the strength of the interaction between the bare states transforms into the level splitting of the dressed states as displayed in \fig{fig:dressed}.
The Hamiltonian $\hbar \nu_j a^\dagger_j a_j$ describing the energy of the vibrational modes is invariant under all these transformations. As a consequence,  the resulting  Hamiltonian 
\BE
\begin{split}
H_\text{dressed}=&\frac{\mu B_j}{2} \sigma_z^{(\pm)} +\hbar \nu_j a^\dagger_j a_j\\
&+ \frac{\mu B' b_{j,n}q_j}{2} (a^\dagger_j +  a_j)\sigma_z^{(\pm)}
\end{split}
\label{eq:H_dressed}
\EE
exhibits exactly the same form as the Hamiltonian of the static gradient field given in \eq{eq:H_static}. A comparison of both Hamiltonians leads to the identifications
\BE
\omega(z)=\frac{\mu B_j}{\hbar}
\EE
and
\BE
\varepsilon_{j,n}=\frac{\mu B' b_{j,n} q_j}{2\hbar \nu_j}\label{eq:coupling} \ .
\EE
Note that using \eq{eq:Omega_nj} we can write $\varepsilon_{j,n}=\Omega_{j,n}/\nu_j$. 

{\em Discussion.}
The scheme proposed here -- dynamical MAGIC combined with dressed states -- does not require to null the dynamic magnetic field at the ion position. Single-qubit gates using the dressed states $\ket{\pm}$ as a qubit can be carried out by employing a resonant RF-field \cite{Timoney2011,Webster2013,Randall2015}. In order to implement conditional quantum gates, for example 2-qubit gates, the application of an RF-field tuned (close) to resonance with a motional sideband transition between dressed states can be used \cite{Cirac1995,Molmer2000,Cohen2015,Weidt2016}. In this case, the effective Lamb-Dicke parameter, $\varepsilon_{j,n}$ allows for the necessary coupling between qubit states and motional states \cite{Mintert2001}. Gates with dressed states in a {\em static} field gradient have been proposed and successfully implemented  with high fidelity \cite{Timoney2011,Cohen2015,Randall2015,Weidt2016}.  

As a concrete example for dynamic MAGIC with dressed states we consider  $^{171}$Yb$^+$ ions exposed to a gradient  of $B'=65$ T/m,  $\nu_1=2\pi \times 500$ kHz,  and a Rabi frequency characterizing the RF gate field  of $2\pi\times 0.1\ \text{MHz}$: according to \cite{Cohen2015} (where dressed states in a static gradient are considered),  we expect a gate time of $200\  \mu\text{s}$  and a gate fidelity in the regime of $0.998$. 

In existing implementations of dynamical MAGIC, a static bias magnetic field having a well defined magnitude is applied to a qubit resonance 
in order to make it only weakly sensitive to ambient magnetic fields, and thus enhance its coherence time \cite{Langer2005,Ospelkaus2011,Lucas2014,Carsjens2014}.  
An additional benefit from using  dressed states that exist in a non-zero dynamic gradient field would be that dressed qubits are already resistant against dephasing by ambient noise fields \cite{Timoney2011,Webster2013,Randall2015,Cohen2015,Weidt2016} without application of an accurately controlled, strong bias field. The dressed state qubit's coherence time would be sensitive to fluctuations in the amplitude of the dressing field. This sensitivity could be strongly reduced by the use of a second dressing field at frequency $\Omega_0$ \cite{Cohen2015}.

Dynamical MAGIC combined with dressed states works with a non-zero dynamic magnetic field and can, therefore, as well be realized with a dynamic gradient along the {\em axial} direction of an  ion string where each ion is exposed to a different non-zero dynamic field. 
Because the axial eigenmodes are characterized by a lower frequency than the radial modes, and the coupling between internal and motional states,  $\epsilon \propto 1/\nu^{-3/2}$, such an arrangement enhances this coupling.

In what follows we consider concrete examples for coupling constants that could be achieved experimentally. Using a dynamic gradient $B'=35$ T/m (as achieved in previous experiments \cite{Ospelkaus2011}) and an axial trap frequency $\nu_1= 300$ kHz, the spin-motion coupling  for two $^9$Be$+$ ions amounts to   $\varepsilon_{j,1}= 0.05$, a magnitude useful for many experiments, for instance, for conditional quantum gates.  
A dynamic gradient $B'=200$ T/m appears realistic in future experiments leading again to $\varepsilon_{j,1} \approx 0.05$, now for an axial trap frequency $\nu_1=2\pi \times 1$ MHz. 

Dressed states created by a dynamic gradient exhibit spin-spin coupling as was shown for the Hamiltonian  \eq{eq:H_static} that describes spin-spin coupling in a static gradient. Importantly, a suitable transformation of Hamiltonian \eq{eq:H_dressed} reveals such a long range spin-spin coupling between all pairs of ions exposed to a dynamic gradient field exactly as given in  \eq{eq:J}. 

The $J$ coupling constant (proportional to $B'^2/\nu_1^2$) in the presence of a gradient $B'=200$ T/m and with $\nu_1=2\pi \times 1$ MHz amounts to $2\pi \times 1.5$ kHz allowing for a CNOT gate time of about 170 $\mu$s. This in turn gives a ratio between experimentally achieved coherence times of dressed states and gate duration of about $3\times 10^4$. So far, $N=2$ ions undergoing conditional quantum dynamics in a trapping zone  have been considered as a building block of a scalable device \cite{Kielpinski2002}. Trapping $N>2$ ions  in a trapping zone and taking advantage of long-range spin-spin coupling opens new possibilities for quantum simulations with individually addressed spins and for multi-qubit gates accelerating quantum algorithms  \cite{Piltz2015}. This long-range spin-spin coupling can be tailored for a specific purpose by adjusting global and local trapping potentials in ion traps, even while carrying out a simulation or computation \cite{McHugh2005,HWunderlich2009,Zippilli2014}. Such on-the-fly tailoring of potentials can be achieved by changing small voltages applied to segmented trap electrodes.

As shown above, the use of states dressed by a dynamic gradient field as qubits removes the experimentally demanding requirement to null the dynamic field at the position of the ions in order to achieve high-fidelity gates. In addition, the application of an axial dynamic gradient leads to stronger coupling between internal and motional states thus allowing for faster quantum gates. Also, spin-spin coupling insensitive to vibrational excitation can be used for quantum simulations and quantum computation.  In fact, a non-zero dynamic field can have a further advantage, since it allows for individual qubit rotations using an RF driving field at frequency $\Omega_0$ specific for each individual ion. Thus, closely spaced, interacting ions exposed to a dynamic gradient could be individually addressed \cite{Navon2013}  in the same way as it was done with a static field gradient by simply dialing in the appropriate RF frequency for each qubit \cite{Johanning2009,Khromova2012,Piltz2014}. Using the parameters given above (two $^9$Be$^+$ ions, $B'=200$ T/m,  $\nu_1=2 \pi\times1$ MHz) we estimate a difference of $\Delta \omega\approx 2\pi\times 10$ MHz in the addressing frequency. The probability $p$ to excite ion A off-resonant with a single-qubit rotation aimed on ion B is determined by $p=\Omega^2/(\Omega^2+\Delta \omega^2)\approx 10^{-4}$ assuming a typical resonant Rabi-frequency $\Omega= 2\pi \times 100$ kHz for ion B.

The novel concept introduced here  -- with the features summarized in the introduction and discussed in more detail above -- should allow for a decisive reduction of experimental complexity and, at the same time, opens new perspectives for an RF-based approach to quantum computation and quantum simulations with trapped ions.

We acknowledge financial support from the Deutsche Forschungsgemeinschaft.

\bibliographystyle{apsrev}

\begin{thebibliography}{40}
\expandafter\ifx\csname natexlab\endcsname\relax\def\natexlab#1{#1}\fi
\expandafter\ifx\csname bibnamefont\endcsname\relax
  \def\bibnamefont#1{#1}\fi
\expandafter\ifx\csname bibfnamefont\endcsname\relax
  \def\bibfnamefont#1{#1}\fi
\expandafter\ifx\csname citenamefont\endcsname\relax
  \def\citenamefont#1{#1}\fi
\expandafter\ifx\csname url\endcsname\relax
  \def\url#1{\texttt{#1}}\fi
\expandafter\ifx\csname urlprefix\endcsname\relax\def\urlprefix{URL }\fi
\providecommand{\bibinfo}[2]{#2}
\providecommand{\eprint}[2][]{\url{#2}}

\bibitem[{\citenamefont{Cirac and Zoller}(1995)}]{Cirac1995}
\bibinfo{author}{\bibfnamefont{J. I.}~\bibnamefont{Cirac}} \bibnamefont{and}
  \bibinfo{author}{\bibfnamefont{P.}~\bibnamefont{Zoller}},
  \bibinfo{journal}{Phys. Rev. Lett.} \textbf{\bibinfo{volume}{74}},
  \bibinfo{pages}{4091} (\bibinfo{year}{1995}).

\bibitem[{\citenamefont{Blatt and Wineland}(2008)}]{Blatt2008}
\bibinfo{author}{\bibfnamefont{R.}~\bibnamefont{Blatt}} \bibnamefont{and}
  \bibinfo{author}{\bibfnamefont{D.}~\bibnamefont{Wineland}},
  \bibinfo{journal}{Nature} \textbf{\bibinfo{volume}{453}},
  \bibinfo{pages}{1008} (\bibinfo{year}{2008}).

\bibitem[{\citenamefont{Wineland}(2013)}]{Wineland2013}
\bibinfo{author}{\bibfnamefont{D.~J.} \bibnamefont{Wineland}},
  \bibinfo{journal}{Reviews of Modern Physics} \textbf{\bibinfo{volume}{85}},
  \bibinfo{pages}{1103} (\bibinfo{year}{2013}).

\bibitem[{\citenamefont{Ladd et~al.}(2010)\citenamefont{Ladd, Jelezko,
  Laflamme, Nakamura, Monroe, and O'Brian}}]{Monroe2010}
\bibinfo{author}{\bibfnamefont{T.}~\bibnamefont{Ladd}},
  \bibinfo{author}{\bibfnamefont{F.}~\bibnamefont{Jelezko}},
  \bibinfo{author}{\bibfnamefont{R.}~\bibnamefont{Laflamme}},
  \bibinfo{author}{\bibfnamefont{Y.}~\bibnamefont{Nakamura}},
  \bibinfo{author}{\bibfnamefont{C.}~\bibnamefont{Monroe}}, \bibnamefont{and}
  \bibinfo{author}{\bibfnamefont{J.}~\bibnamefont{O'Brian}},
  \bibinfo{journal}{Nature} \textbf{\bibinfo{volume}{464}}, \bibinfo{pages}{45}
  (\bibinfo{year}{2010}).

\bibitem[{\citenamefont{Stenholm}(1986)}]{Stenholm1986}
\bibinfo{author}{\bibfnamefont{S.}~\bibnamefont{Stenholm}},
  \bibinfo{journal}{Rev. Mo. Phys.} \textbf{\bibinfo{volume}{58}},
  \bibinfo{pages}{699} (\bibinfo{year}{1986}).

\bibitem[{\citenamefont{Hanneke et~al.}(2010)\citenamefont{Hanneke, Home, Jost,
  Amini, Leibfried, and Wineland}}]{Hanneke2010}
\bibinfo{author}{\bibfnamefont{D.}~\bibnamefont{Hanneke}},
  \bibinfo{author}{\bibfnamefont{J.}~\bibnamefont{Home}},
  \bibinfo{author}{\bibfnamefont{J.}~\bibnamefont{Jost}},
  \bibinfo{author}{\bibfnamefont{J.}~\bibnamefont{Amini}},
  \bibinfo{author}{\bibfnamefont{D.}~\bibnamefont{Leibfried}},
  \bibnamefont{and} \bibinfo{author}{\bibfnamefont{D.}~\bibnamefont{Wineland}},
  \bibinfo{journal}{Nature Phys.} \textbf{\bibinfo{volume}{6}},
  \bibinfo{pages}{13} (\bibinfo{year}{2010}).

\bibitem[{\citenamefont{Monz et~al.}(2016)\citenamefont{Monz, Nigg, Martinez,
  Brandl, Schindler, Rines, Wang, Chuang, and Blatt}}]{Blatt2016}
\bibinfo{author}{\bibfnamefont{T.}~\bibnamefont{Monz}},
  \bibinfo{author}{\bibfnamefont{D.}~\bibnamefont{Nigg}},
  \bibinfo{author}{\bibfnamefont{E.}~\bibnamefont{Martinez}},
  \bibinfo{author}{\bibfnamefont{F.~M.} \bibnamefont{Brandl}},
  \bibinfo{author}{\bibfnamefont{P.}~\bibnamefont{Schindler}},
  \bibinfo{author}{\bibfnamefont{R.}~\bibnamefont{Rines}},
  \bibinfo{author}{\bibfnamefont{S.~X.} \bibnamefont{Wang}},
  \bibinfo{author}{\bibfnamefont{I.}~\bibnamefont{Chuang}}, \bibnamefont{and}
  \bibinfo{author}{\bibfnamefont{R.}~\bibnamefont{Blatt}},
  \bibinfo{journal}{Science} \textbf{\bibinfo{volume}{351}},
  \bibinfo{pages}{1068} (\bibinfo{year}{2016}).

\bibitem[{\citenamefont{Schneider et~al.}(2012)\citenamefont{Schneider, Porras,
  and Schaetz}}]{Schneider2012}
\bibinfo{author}{\bibfnamefont{C.}~\bibnamefont{Schneider}},
  \bibinfo{author}{\bibfnamefont{D.}~\bibnamefont{Porras}}, \bibnamefont{and}
  \bibinfo{author}{\bibfnamefont{T.}~\bibnamefont{Schaetz}},
  \bibinfo{journal}{Rep. Prog. Phys.} \textbf{\bibinfo{volume}{75}},
  \bibinfo{pages}{024401} (\bibinfo{year}{2012}).

\bibitem[{\citenamefont{Islam et~al.}(2013)\citenamefont{Islam, Senko,
  Campbell, Korenblit, Smith, Lee, Edwards, Wang, Freericks, and
  Monroe}}]{Islam2013}
\bibinfo{author}{\bibfnamefont{R.}~\bibnamefont{Islam}},
  \bibinfo{author}{\bibfnamefont{C.}~\bibnamefont{Senko}},
  \bibinfo{author}{\bibfnamefont{W.~C.} \bibnamefont{Campbell}},
  \bibinfo{author}{\bibfnamefont{S.}~\bibnamefont{Korenblit}},
  \bibinfo{author}{\bibfnamefont{J.}~\bibnamefont{Smith}},
  \bibinfo{author}{\bibfnamefont{A.}~\bibnamefont{Lee}},
  \bibinfo{author}{\bibfnamefont{E.~E.} \bibnamefont{Edwards}},
  \bibinfo{author}{\bibfnamefont{C.-C.~J.} \bibnamefont{Wang}},
  \bibinfo{author}{\bibfnamefont{J.~K.} \bibnamefont{Freericks}},
  \bibnamefont{and} \bibinfo{author}{\bibfnamefont{C.}~\bibnamefont{Monroe}},
  \bibinfo{journal}{Science} \textbf{\bibinfo{volume}{340}},
  \bibinfo{pages}{583} (\bibinfo{year}{2013}).

\bibitem[{\citenamefont{Mintert and Wunderlich}(2001)}]{Mintert2001}
\bibinfo{author}{\bibfnamefont{F.}~\bibnamefont{Mintert}} \bibnamefont{and}
  \bibinfo{author}{\bibfnamefont{C.}~\bibnamefont{Wunderlich}},
  \bibinfo{journal}{Phys. Rev. Lett.} \textbf{\bibinfo{volume}{87}},
  \bibinfo{pages}{257904} (\bibinfo{year}{2001}).

\bibitem[{\citenamefont{Ospelkaus et~al.}(2008)\citenamefont{Ospelkaus, Langer,
  Amini, Brown, Leibfried, and Wineland}}]{Ospelkaus2008}
\bibinfo{author}{\bibfnamefont{C.}~\bibnamefont{Ospelkaus}},
  \bibinfo{author}{\bibfnamefont{C.~E.} \bibnamefont{Langer}},
  \bibinfo{author}{\bibfnamefont{J.-M.} \bibnamefont{Amini}},
  \bibinfo{author}{\bibfnamefont{K.~R.} \bibnamefont{Brown}},
  \bibinfo{author}{\bibfnamefont{D.}~\bibnamefont{Leibfried}},
  \bibnamefont{and} \bibinfo{author}{\bibfnamefont{D.~J.}
  \bibnamefont{Wineland}}, \bibinfo{journal}{Phys. Rev. Lett.}
  \textbf{\bibinfo{volume}{101}}, \bibinfo{pages}{090502}
  (\bibinfo{year}{2008}).

\bibitem[{\citenamefont{Johanning et~al.}(2009)\citenamefont{Johanning, Braun,
  Timoney, Elman, Neuhauser, and Wunderlich}}]{Johanning2009}
\bibinfo{author}{\bibfnamefont{M.}~\bibnamefont{Johanning}},
  \bibinfo{author}{\bibfnamefont{A.}~\bibnamefont{Braun}},
  \bibinfo{author}{\bibfnamefont{N.}~\bibnamefont{Timoney}},
  \bibinfo{author}{\bibfnamefont{V.}~\bibnamefont{Elman}},
  \bibinfo{author}{\bibfnamefont{W.}~\bibnamefont{Neuhauser}},
  \bibnamefont{and}
  \bibinfo{author}{\bibfnamefont{C.}~\bibnamefont{Wunderlich}},
  \bibinfo{journal}{Phys. Rev. Lett.} \textbf{\bibinfo{volume}{102}},
  \bibinfo{pages}{073004} (\bibinfo{year}{2009}).

\bibitem[{\citenamefont{Ospelkaus et~al.}(2011)\citenamefont{Ospelkaus,
  Warring, Colombe, Brown, Amini, Leibfried, and Wineland}}]{Ospelkaus2011}
\bibinfo{author}{\bibfnamefont{C.}~\bibnamefont{Ospelkaus}},
  \bibinfo{author}{\bibfnamefont{U.}~\bibnamefont{Warring}},
  \bibinfo{author}{\bibfnamefont{Y.}~\bibnamefont{Colombe}},
  \bibinfo{author}{\bibfnamefont{K.}~\bibnamefont{Brown}},
  \bibinfo{author}{\bibfnamefont{J.}~\bibnamefont{Amini}},
  \bibinfo{author}{\bibfnamefont{D.}~\bibnamefont{Leibfried}},
  \bibnamefont{and} \bibinfo{author}{\bibfnamefont{D.}~\bibnamefont{Wineland}},
  \bibinfo{journal}{Nature} \textbf{\bibinfo{volume}{476}},
  \bibinfo{pages}{181} (\bibinfo{year}{2011}).

\bibitem[{\citenamefont{Khromova et~al.}(2012)\citenamefont{Khromova, Piltz,
  Scharfenberger, Gloger, Johanning, Var\'on, and Wunderlich}}]{Khromova2012}
\bibinfo{author}{\bibfnamefont{A.}~\bibnamefont{Khromova}},
  \bibinfo{author}{\bibfnamefont{C.}~\bibnamefont{Piltz}},
  \bibinfo{author}{\bibfnamefont{B.}~\bibnamefont{Scharfenberger}},
  \bibinfo{author}{\bibfnamefont{T.~F.} \bibnamefont{Gloger}},
  \bibinfo{author}{\bibfnamefont{M.}~\bibnamefont{Johanning}},
  \bibinfo{author}{\bibfnamefont{A.~F.} \bibnamefont{Var\'on}},
  \bibnamefont{and}
  \bibinfo{author}{\bibfnamefont{C.}~\bibnamefont{Wunderlich}},
  \bibinfo{journal}{Phys. Rev. Lett.} \textbf{\bibinfo{volume}{108}},
  \bibinfo{pages}{220502} (\bibinfo{year}{2012}).

\bibitem[{\citenamefont{Belmechri et~al.}(2013)\citenamefont{Belmechri,
  F\"orster, Alt, Widera, Meschede, and Alberti}}]{Belmechri2013}
\bibinfo{author}{\bibfnamefont{N.}~\bibnamefont{Belmechri}},
  \bibinfo{author}{\bibfnamefont{L.}~\bibnamefont{F\"orster}},
  \bibinfo{author}{\bibfnamefont{W.}~\bibnamefont{Alt}},
  \bibinfo{author}{\bibfnamefont{A.}~\bibnamefont{Widera}},
  \bibinfo{author}{\bibfnamefont{D.}~\bibnamefont{Meschede}}, \bibnamefont{and}
  \bibinfo{author}{\bibfnamefont{A.}~\bibnamefont{Alberti}},
  \bibinfo{journal}{J. Phys. B.} \textbf{\bibinfo{volume}{46}},
  \bibinfo{pages}{104006} (\bibinfo{year}{2013}).

\bibitem[{\citenamefont{Lake et~al.}(2015)\citenamefont{Lake, Weidt, Randall,
  Standing, Webster, and Hensinger}}]{Lake2015}
\bibinfo{author}{\bibfnamefont{K.}~\bibnamefont{Lake}},
  \bibinfo{author}{\bibfnamefont{S.}~\bibnamefont{Weidt}},
  \bibinfo{author}{\bibfnamefont{J.}~\bibnamefont{Randall}},
  \bibinfo{author}{\bibfnamefont{E. D.}~\bibnamefont{Standing}},
  \bibinfo{author}{\bibfnamefont{S. C.}~\bibnamefont{Webster}}, \bibnamefont{and}
  \bibinfo{author}{\bibfnamefont{W. K.}~\bibnamefont{Hensinger}},
  \bibinfo{journal}{Phys. Rev. A} \textbf{\bibinfo{volume}{91}},
  \bibinfo{pages}{012319} (\bibinfo{year}{2015}).

\bibitem[{\citenamefont{Weidt et~al.}()\citenamefont{Weidt, Randall, Webster,
  Lake, Webb, Cohen, Navickas, Lekitsch, Retzker, and Hensinger}}]{Weidt2016}
\bibinfo{author}{\bibfnamefont{S.}~\bibnamefont{Weidt}},
  \bibinfo{author}{\bibfnamefont{J.}~\bibnamefont{Randall}},
  \bibinfo{author}{\bibfnamefont{S.~C.} \bibnamefont{Webster}},
  \bibinfo{author}{\bibfnamefont{K.}~\bibnamefont{Lake}},
  \bibinfo{author}{\bibfnamefont{A.~E.} \bibnamefont{Webb}},
  \bibinfo{author}{\bibfnamefont{I.}~\bibnamefont{Cohen}},
  \bibinfo{author}{\bibfnamefont{T.}~\bibnamefont{Navickas}},
  \bibinfo{author}{\bibfnamefont{B.}~\bibnamefont{Lekitsch}},
  \bibinfo{author}{\bibfnamefont{A.}~\bibnamefont{Retzker}}, \bibnamefont{and}
  \bibinfo{author}{\bibfnamefont{W.~K.} \bibnamefont{Hensinger}},
  \bibinfo{note}{arXiv:1603.03384}.

\bibitem[{\citenamefont{Schrader et~al.}(2004)\citenamefont{Schrader, Dotsenko,
  Khudaverdyan, Miroshnychenko, Rauschenbeutel, and Meschede}}]{Schrader2004}
\bibinfo{author}{\bibfnamefont{D.}~\bibnamefont{Schrader}},
  \bibinfo{author}{\bibfnamefont{I.}~\bibnamefont{Dotsenko}},
  \bibinfo{author}{\bibfnamefont{M.}~\bibnamefont{Khudaverdyan}},
  \bibinfo{author}{\bibfnamefont{Y.}~\bibnamefont{Miroshnychenko}},
  \bibinfo{author}{\bibfnamefont{A.}~\bibnamefont{Rauschenbeutel}},
  \bibnamefont{and} \bibinfo{author}{\bibfnamefont{D.}~\bibnamefont{Meschede}},
  \bibinfo{journal}{Phys. Rev. Lett.} \textbf{\bibinfo{volume}{93}},
  \bibinfo{pages}{150501} (\bibinfo{year}{2004}).

\bibitem[{\citenamefont{Warring
  et~al.}(2013{\natexlab{a}})\citenamefont{Warring, Ospelkaus, Colombe,
  J\"ordens, Leibfried, and Wineland}}]{Warring2013}
\bibinfo{author}{\bibfnamefont{U.}~\bibnamefont{Warring}},
  \bibinfo{author}{\bibfnamefont{C.}~\bibnamefont{Ospelkaus}},
  \bibinfo{author}{\bibfnamefont{Y.}~\bibnamefont{Colombe}},
  \bibinfo{author}{\bibfnamefont{R.}~\bibnamefont{J\"ordens}},
  \bibinfo{author}{\bibfnamefont{D.}~\bibnamefont{Leibfried}},
  \bibnamefont{and} \bibinfo{author}{\bibfnamefont{D. J.}~\bibnamefont{Wineland}},
  \bibinfo{journal}{Phys. Rev. Lett.} \textbf{\bibinfo{volume}{110}},
  \bibinfo{pages}{173002} (\bibinfo{year}{2013}{\natexlab{a}}).

\bibitem[{\citenamefont{Piltz et~al.}(2014)\citenamefont{Piltz, Sriarunothai,
  Var\'on, and Wunderlich}}]{Piltz2014}
\bibinfo{author}{\bibfnamefont{C.}~\bibnamefont{Piltz}},
  \bibinfo{author}{\bibfnamefont{T.}~\bibnamefont{Sriarunothai}},
  \bibinfo{author}{\bibfnamefont{A.}~\bibnamefont{Var\'on}}, \bibnamefont{and}
  \bibinfo{author}{\bibfnamefont{C.}~\bibnamefont{Wunderlich}},
  \bibinfo{journal}{Nat. Commun.} \textbf{\bibinfo{volume}{5}},
  \bibinfo{pages}{679} (\bibinfo{year}{2014}).

\bibitem[{\citenamefont{Aude~Craik et~al.}()\citenamefont{Aude~Craik, Linke,
  Sepiol, Harty, Ballance, Stacey, Steane, Lucas, and Allcock}}]{Aude2016}
\bibinfo{author}{\bibfnamefont{D.}~\bibnamefont{Aude~Craik}},
  \bibinfo{author}{\bibfnamefont{N.}~\bibnamefont{Linke}},
  \bibinfo{author}{\bibfnamefont{M.}~\bibnamefont{Sepiol}},
  \bibinfo{author}{\bibfnamefont{T. P.}~\bibnamefont{Harty}},
  \bibinfo{author}{\bibfnamefont{C.}~\bibnamefont{Ballance}},
  \bibinfo{author}{\bibfnamefont{D.}~\bibnamefont{Stacey}},
  \bibinfo{author}{\bibfnamefont{A.}~\bibnamefont{Steane}},
  \bibinfo{author}{\bibfnamefont{D.}~\bibnamefont{Lucas}}, \bibnamefont{and}
  \bibinfo{author}{\bibfnamefont{D.}~\bibnamefont{Allcock}},
  \bibinfo{note}{arXiv:1601.02696}.

\bibitem[{\citenamefont{Brown et~al.}(2011)\citenamefont{Brown, Wilson,
  Colombe, Ospelkaus, Meier, Knill, Leibfried, and Wineland}}]{Brown2011}
\bibinfo{author}{\bibfnamefont{K.~R.} \bibnamefont{Brown}},
  \bibinfo{author}{\bibfnamefont{A.~C.} \bibnamefont{Wilson}},
  \bibinfo{author}{\bibfnamefont{Y.}~\bibnamefont{Colombe}},
  \bibinfo{author}{\bibfnamefont{C.}~\bibnamefont{Ospelkaus}},
  \bibinfo{author}{\bibfnamefont{A.~M.} \bibnamefont{Meier}},
  \bibinfo{author}{\bibfnamefont{E.}~\bibnamefont{Knill}},
  \bibinfo{author}{\bibfnamefont{D.}~\bibnamefont{Leibfried}},
  \bibnamefont{and} \bibinfo{author}{\bibfnamefont{D.~J.}
  \bibnamefont{Wineland}}, \bibinfo{journal}{Phys. Rev. A}
  \textbf{\bibinfo{volume}{84}}, \bibinfo{pages}{030303(R)}
  (\bibinfo{year}{2011}).

\bibitem[{\citenamefont{Harty et~al.}(2014)\citenamefont{Harty, Allcock,
  Ballance, Guidoni, Janacek, Linke, Stacey, and Lucas}}]{Lucas2014}
\bibinfo{author}{\bibfnamefont{T. P.}~\bibnamefont{Harty}},
  \bibinfo{author}{\bibfnamefont{D. T. C.}~\bibnamefont{Allcock}},
  \bibinfo{author}{\bibfnamefont{C. J.}~\bibnamefont{Ballance}},
  \bibinfo{author}{\bibfnamefont{L.}~\bibnamefont{Guidoni}},
  \bibinfo{author}{\bibfnamefont{H. A.}~\bibnamefont{Janacek}},
  \bibinfo{author}{\bibfnamefont{N. M.}~\bibnamefont{Linke}},
  \bibinfo{author}{\bibfnamefont{D. N.}~\bibnamefont{Stacey}}, \bibnamefont{and}
  \bibinfo{author}{\bibfnamefont{D. M.}~\bibnamefont{Lucas}},
  \bibinfo{journal}{Phys. Rev. Lett.} \textbf{\bibinfo{volume}{113}},
  \bibinfo{pages}{220501} (\bibinfo{year}{2014}).

\bibitem[{\citenamefont{Piltz et~al.}()\citenamefont{Piltz, Sriarunothai,
  Ivanov, W\"olk, and Wunderlich}}]{Piltz2015}
\bibinfo{author}{\bibfnamefont{C.}~\bibnamefont{Piltz}},
  \bibinfo{author}{\bibfnamefont{T.}~\bibnamefont{Sriarunothai}},
  \bibinfo{author}{\bibfnamefont{S.}~\bibnamefont{Ivanov}},
  \bibinfo{author}{\bibfnamefont{S.}~\bibnamefont{W\"olk}}, \bibnamefont{and}
  \bibinfo{author}{\bibfnamefont{C.}~\bibnamefont{Wunderlich}},
  \bibinfo{note}{arXiv: 1509.01478}.

\bibitem[{\citenamefont{Johanning and Wunderlich}(2009)}]{Johanning2009b}
\bibinfo{author}{\bibfnamefont{A.}~\bibnamefont{Johanning},
  \bibfnamefont{M.~Varon}} \bibnamefont{and}
  \bibinfo{author}{\bibfnamefont{C.}~\bibnamefont{Wunderlich}},
  \bibinfo{journal}{J. Phys. B.} \textbf{\bibinfo{volume}{42}},
  \bibinfo{pages}{154009} (\bibinfo{year}{2009}).

\bibitem[{\citenamefont{Warring
  et~al.}(2013{\natexlab{b}})\citenamefont{Warring, Ospelkaus, Colombe, Brown,
  Amini, Carsjens, Leibfried, and Wineland}}]{Warring2013b}
\bibinfo{author}{\bibfnamefont{U.}~\bibnamefont{Warring}},
  \bibinfo{author}{\bibfnamefont{C.}~\bibnamefont{Ospelkaus}},
  \bibinfo{author}{\bibfnamefont{Y.}~\bibnamefont{Colombe}},
  \bibinfo{author}{\bibfnamefont{K.~R.} \bibnamefont{Brown}},
  \bibinfo{author}{\bibfnamefont{J.~M.} \bibnamefont{Amini}},
  \bibinfo{author}{\bibfnamefont{M.}~\bibnamefont{Carsjens}},
  \bibinfo{author}{\bibfnamefont{D.}~\bibnamefont{Leibfried}},
  \bibnamefont{and} \bibinfo{author}{\bibfnamefont{D.~J.}
  \bibnamefont{Wineland}}, \bibinfo{journal}{Phys. Rev. A}
  \textbf{\bibinfo{volume}{87}}, \bibinfo{pages}{013437}
  (\bibinfo{year}{2013}{\natexlab{b}}).

\bibitem[{\citenamefont{Carsjens et~al.}(2014)\citenamefont{Carsjens, Kohnen,
  Dubielzig, and Ospelkaus}}]{Carsjens2014}
\bibinfo{author}{\bibfnamefont{M.}~\bibnamefont{Carsjens}},
  \bibinfo{author}{\bibfnamefont{M.}~\bibnamefont{Kohnen}},
  \bibinfo{author}{\bibfnamefont{T.}~\bibnamefont{Dubielzig}},
  \bibnamefont{and}
  \bibinfo{author}{\bibfnamefont{C.}~\bibnamefont{Ospelkaus}},
  \bibinfo{journal}{Appl. Phys. B} \textbf{\bibinfo{volume}{114}},
  \bibinfo{pages}{243} (\bibinfo{year}{2014}).

\bibitem[{\citenamefont{Wunderlich}(2002)}]{Wunderlich2002}
\bibinfo{author}{\bibfnamefont{C.}~\bibnamefont{Wunderlich}},
  \emph{\bibinfo{title}{Conditional Spin Resonance with Trapped Ions}}
  (\bibinfo{publisher}{Springer}, \bibinfo{address}{Berlin},
  \bibinfo{year}{2002}), p. \bibinfo{pages}{261}.

\bibitem[{\citenamefont{Wunderlich and Balzer}(2003)}]{Wunderlich2003}
\bibinfo{author}{\bibfnamefont{C.}~\bibnamefont{Wunderlich}} \bibnamefont{and}
  \bibinfo{author}{\bibfnamefont{C.}~\bibnamefont{Balzer}},
  \bibinfo{journal}{Adv. At. Mol. Opt. Phys.} \textbf{\bibinfo{volume}{49}},
  \bibinfo{pages}{293} (\bibinfo{year}{2003}).

\bibitem[{\citenamefont{Timoney et~al.}(2011)\citenamefont{Timoney, Baumgart,
  Johanning, Var\'on, Plenio, Retzker, and Wunderlich}}]{Timoney2011}
\bibinfo{author}{\bibfnamefont{N.}~\bibnamefont{Timoney}},
  \bibinfo{author}{\bibfnamefont{I.}~\bibnamefont{Baumgart}},
  \bibinfo{author}{\bibfnamefont{M.}~\bibnamefont{Johanning}},
  \bibinfo{author}{\bibfnamefont{A.}~\bibnamefont{Var\'on}},
  \bibinfo{author}{\bibfnamefont{M.}~\bibnamefont{Plenio}},
  \bibinfo{author}{\bibfnamefont{A.}~\bibnamefont{Retzker}}, \bibnamefont{and}
  \bibinfo{author}{\bibfnamefont{C.}~\bibnamefont{Wunderlich}},
  \bibinfo{journal}{Nature} \textbf{\bibinfo{volume}{476}},
  \bibinfo{pages}{185} (\bibinfo{year}{2011}).

\bibitem[{\citenamefont{Webster et~al.}(2013)\citenamefont{Webster, Weidt,
  Lake, McLoughlin, and Hensinger}}]{Webster2013}
\bibinfo{author}{\bibfnamefont{S. C.}~\bibnamefont{Webster}},
  \bibinfo{author}{\bibfnamefont{S.}~\bibnamefont{Weidt}},
  \bibinfo{author}{\bibfnamefont{K.}~\bibnamefont{Lake}},
  \bibinfo{author}{\bibfnamefont{J. J.}~\bibnamefont{McLoughlin}},
  \bibnamefont{and}
  \bibinfo{author}{\bibfnamefont{W. K.}~\bibnamefont{Hensinger}},
  \bibinfo{journal}{Phys. Rev. Lett.} \textbf{\bibinfo{volume}{111}},
  \bibinfo{pages}{140501} (\bibinfo{year}{2013}).

\bibitem[{\citenamefont{Randall et~al.}(2015)\citenamefont{Randall, Weidt,
  Standing, Lake, Webster, Murgia, Navickas, Roth, and
  Hensinger}}]{Randall2015}
\bibinfo{author}{\bibfnamefont{J.}~\bibnamefont{Randall}},
  \bibinfo{author}{\bibfnamefont{S.}~\bibnamefont{Weidt}},
  \bibinfo{author}{\bibfnamefont{E.}~\bibnamefont{Standing}},
  \bibinfo{author}{\bibfnamefont{K.}~\bibnamefont{Lake}},
  \bibinfo{author}{\bibfnamefont{S.}~\bibnamefont{Webster}},
  \bibinfo{author}{\bibfnamefont{D.}~\bibnamefont{Murgia}},
  \bibinfo{author}{\bibfnamefont{T.}~\bibnamefont{Navickas}},
  \bibinfo{author}{\bibfnamefont{K.}~\bibnamefont{Roth}}, \bibnamefont{and}
  \bibinfo{author}{\bibfnamefont{W.}~\bibnamefont{Hensinger}},
  \bibinfo{journal}{Phys. Rev. A} \textbf{\bibinfo{volume}{91}},
  \bibinfo{pages}{012319} (\bibinfo{year}{2015}).

\bibitem[{\citenamefont{S\o~rensen and M\o~lmer}(2000)}]{Molmer2000}
\bibinfo{author}{\bibfnamefont{A.}~\bibnamefont{S\o~rensen}} \bibnamefont{and}
  \bibinfo{author}{\bibfnamefont{K.}~\bibnamefont{M\o~lmer}},
  \bibinfo{journal}{Phys. Rev. A} \textbf{\bibinfo{volume}{62}},
  \bibinfo{pages}{022311} (\bibinfo{year}{2000}).

\bibitem[{\citenamefont{Cohen et~al.}(2015)\citenamefont{Cohen, Weidt,
  Hensinger, and Retzker}}]{Cohen2015}
\bibinfo{author}{\bibfnamefont{I.}~\bibnamefont{Cohen}},
  \bibinfo{author}{\bibfnamefont{S.}~\bibnamefont{Weidt}},
  \bibinfo{author}{\bibfnamefont{W.}~\bibnamefont{Hensinger}},
  \bibnamefont{and} \bibinfo{author}{\bibfnamefont{A.}~\bibnamefont{Retzker}},
  \bibinfo{journal}{New. J. Phys.} \textbf{\bibinfo{volume}{17}},
  \bibinfo{pages}{043008} (\bibinfo{year}{2015}).

\bibitem[{\citenamefont{Langer et~al.}(2005)\citenamefont{Langer, Ozeri, Jost,
  Chiaverini, DeMarco, Ben-Kish, Blakestad, Britton, Hume, Itano
  et~al.}}]{Langer2005}
\bibinfo{author}{\bibfnamefont{C.}~\bibnamefont{Langer}},
  \bibinfo{author}{\bibfnamefont{R.}~\bibnamefont{Ozeri}},
  \bibinfo{author}{\bibfnamefont{J.~D.} \bibnamefont{Jost}},
  \bibinfo{author}{\bibfnamefont{J.}~\bibnamefont{Chiaverini}},
  \bibinfo{author}{\bibfnamefont{B.}~\bibnamefont{DeMarco}},
  \bibinfo{author}{\bibfnamefont{A.}~\bibnamefont{Ben-Kish}},
  \bibinfo{author}{\bibfnamefont{R.~B.} \bibnamefont{Blakestad}},
  \bibinfo{author}{\bibfnamefont{J.}~\bibnamefont{Britton}},
  \bibinfo{author}{\bibfnamefont{D.~B.} \bibnamefont{Hume}},
  \bibinfo{author}{\bibfnamefont{W.~M.} \bibnamefont{Itano}},
  \bibnamefont{et~al.}, \bibinfo{journal}{Phys. Rev. Lett.}
  \textbf{\bibinfo{volume}{95}}, \bibinfo{pages}{060502}
  (\bibinfo{year}{2005}).

\bibitem[{\citenamefont{Kielpinski et~al.}(2002)\citenamefont{Kielpinski,
  Monroe, and Wineland}}]{Kielpinski2002}
\bibinfo{author}{\bibfnamefont{D.}~\bibnamefont{Kielpinski}},
  \bibinfo{author}{\bibfnamefont{C.}~\bibnamefont{Monroe}}, \bibnamefont{and}
  \bibinfo{author}{\bibfnamefont{D.~J.} \bibnamefont{Wineland}},
  \bibinfo{journal}{Nature} \textbf{\bibinfo{volume}{417}},
  \bibinfo{pages}{709} (\bibinfo{year}{2002}),
  \bibinfo{note}{10.1038/nature00784}.

\bibitem[{\citenamefont{Hugh and Twamley}(2005)}]{McHugh2005}
\bibinfo{author}{\bibfnamefont{D.} \bibnamefont{McHugh}} \bibnamefont{and}
  \bibinfo{author}{\bibfnamefont{J.}~\bibnamefont{Twamley}},
  \bibinfo{journal}{Phys. Rev. A} \textbf{\bibinfo{volume}{71}},
  \bibinfo{pages}{012315} (\bibinfo{year}{2005}).

\bibitem[{\citenamefont{Wunderlich et~al.}(2009)\citenamefont{Wunderlich,
  Wunderlich, Singer, and Schmidt-Kaler}}]{HWunderlich2009}
\bibinfo{author}{\bibfnamefont{H.}~\bibnamefont{Wunderlich}},
  \bibinfo{author}{\bibfnamefont{C.}~\bibnamefont{Wunderlich}},
  \bibinfo{author}{\bibfnamefont{K.}~\bibnamefont{Singer}}, \bibnamefont{and}
  \bibinfo{author}{\bibfnamefont{F.}~\bibnamefont{Schmidt-Kaler}},
  \bibinfo{journal}{Phys. Rev. A} \textbf{\bibinfo{volume}{79}},
  \bibinfo{pages}{052324} (\bibinfo{year}{2009}).

\bibitem[{\citenamefont{Zippilli et~al.}(2014)\citenamefont{Zippilli,
  Johanning, Giampaolo, Wunderlich, and Illuminati}}]{Zippilli2014}
\bibinfo{author}{\bibfnamefont{S.}~\bibnamefont{Zippilli}},
  \bibinfo{author}{\bibfnamefont{M.}~\bibnamefont{Johanning}},
  \bibinfo{author}{\bibfnamefont{S.~M.} \bibnamefont{Giampaolo}},
  \bibinfo{author}{\bibfnamefont{C.}~\bibnamefont{Wunderlich}},
  \bibnamefont{and}
  \bibinfo{author}{\bibfnamefont{F.}~\bibnamefont{Illuminati}},
  \bibinfo{journal}{Phys. Rev. A} \textbf{\bibinfo{volume}{89}},
  \bibinfo{pages}{042308} (\bibinfo{year}{2014}).

\bibitem[{\citenamefont{Navon et~al.}(2013)\citenamefont{Navon, Kotler,
  Akerman, Glickman, Almog, and Ozeri}}]{Navon2013}
\bibinfo{author}{\bibfnamefont{N.}~\bibnamefont{Navon}},
  \bibinfo{author}{\bibfnamefont{S.}~\bibnamefont{Kotler}},
  \bibinfo{author}{\bibfnamefont{N.}~\bibnamefont{Akerman}},
  \bibinfo{author}{\bibfnamefont{Y.}~\bibnamefont{Glickman}},
  \bibinfo{author}{\bibfnamefont{I.}~\bibnamefont{Almog}}, \bibnamefont{and}
  \bibinfo{author}{\bibfnamefont{R.}~\bibnamefont{Ozeri}},
  \bibinfo{journal}{Phys. Rev. Lett.} \textbf{\bibinfo{volume}{111}},
  \bibinfo{pages}{073001} (\bibinfo{year}{2013}).

\end{thebibliography}

\end{document}